\documentclass[11pt]{article}
\pdfoutput=1

\usepackage{authblk}
\usepackage{euscript}
\usepackage{amssymb}
\usepackage{amsfonts}
\usepackage{amsbsy}
\usepackage{epsfig}
\usepackage{amsthm}
\usepackage{amscd}
\usepackage{amstext}
\usepackage{verbatim}
\usepackage{amsmath}
\usepackage{cancel}
\usepackage{capt-of}
\usepackage{empheq}
\usepackage{subfigure}
\usepackage{xcolor}

\usepackage{cite}
\usepackage{bm}
\usepackage{makecell}
\usepackage{longtable}
\usepackage{graphicx}

\usepackage{braket}

\usepackage[pdftex,linktoc=all]{hyperref}
\hypersetup{ 
colorlinks=true, 
linkcolor=black, 
citecolor=red, 
}

\def\ben{\begin{equation}}
\def\een{\end{equation}}

\let\w=\omega

\def\be{\begin{equation}}
\def\ee{\end{equation}}
\def\beq{\begin{equation}}
\def\eeq{\end{equation}}
\def\ba{\begin{array}}
\def\ea{\end{array}}

\def\dalemb#1#2{{\vbox{\hrule height .#2pt
       \hbox{\vrule width.#2pt height#1pt \kern#1pt
               \vrule width.#2pt}
       \hrule height.#2pt}}}

\newcommand{\bea}{\begin{eqnarray}}
\newcommand{\eea}{\end{eqnarray}}

\makeatletter
\newcommand*\bigcdot{\mathpalette\bigcdot@{.5}}
\newcommand*\bigcdot@[2]{\mathbin{\vcenter{\hbox{\scalebox{#2}{$\m@th#1\bullet$}}}}}
\makeatother

\renewcommand{\eqref}[1]{(\ref{#1})}

\title{On the Planckian bound for heat diffusion in insulators}
\author[1]{Connie H. Mousatov}
\author[1,2]{Sean A. Hartnoll}
\affil[1]
{\it \small Department of Physics, Stanford University,
Stanford, CA 94305-4060, USA}
\affil[2]{\it \small
Stanford Institute for Materials and Energy Science, SLAC National Accelerator Laboratory, 2575 Sand Hill Road, Menlo Park,
CA 94025, USA}

\date{\today}

\begin{document}

\maketitle

\begin{abstract}
High temperature thermal transport in insulators has been conjectured to be subject to a Planckian bound on the transport lifetime $\tau \gtrsim \tau_\text{Pl} \equiv \hbar/(k_B T)$, despite phonon dynamics being entirely classical at these temperatures. We argue that this Planckian bound is due to a quantum mechanical bound on the sound velocity: $v_s < v_M$. The `melting velocity' $v_M$ is defined in terms of the melting temperature of the crystal, the interatomic spacing and Planck's constant. We show that for several classes of insulating crystals, both simple and complex, $\tau/\tau_\text{Pl} \approx v_M/v_s$ at high temperatures. The velocity bound therefore implies the Planckian bound.
\end{abstract}

\newpage

\subsection*{Introduction}

It has recently been pointed out \cite{PhysRevLett.120.125901, KamranAharon, KA2} that thermal transport in insulating crystals is consistent with a `Planckian bound' \cite{jan} on the transport lifetime
\be\label{eq:planck}
\frac{1}{\tau} \lesssim \frac{1}{\tau_\text{Pl}} \equiv \frac{k_B T}{\hbar} \,. 
\ee
In \cite{PhysRevLett.120.125901, KamranAharon, KA2} the lifetime is defined by writing the thermal diffusivity $D = v_s^2 \tau$ where $v_s$ is the sound velocity, in the spirit of \cite{Hartnoll2014}. It was noted that for the IV-VI semiconductors 
PbTe, PbSe and PbS, $\tau \sim 6 \, \tau_\text{Pl}$ at high temperatures and that for several perovskites at high temperatures, $\tau \sim 1 - 3 \, \tau_\text{Pl}$. We will consider many more compounds in Fig. \ref{fig:ratios} below.

The importance of the timescale $\hbar/(k_B T)$ in many-body physics has been appreciated for a long time, e.g. \cite{P1, P2, subir}. Recent interest has been ignited by the observation that this timescale controls transport in unconventional metals across diverse parameter regimes \cite{Bruin804, Zhang5378, 2018arXiv180807564Z, Legros2019, graphene}. The appearance of Planckian transport in insulating phonon systems offers a simpler arena in which Planckian dynamics can be probed. Indeed, the physics of thermal conduction due to anharmonic phonons in insulators has been succesfully reproduced at a quantitative level from ab initio computations. See e.g. \cite{Feng_2015,PhysRevB.92.054301} for the Planckian transport regime of SrTiO$_3$ and \cite{slack} for a review of early computations.
Our objective here is instead to understand the physical origin of the quantum mechanical constraint (\ref{eq:planck}) on the phonon dynamics, that would otherwise appear to be in a deeply classical, high temperature regime.

The simple observation that we shall make is inspired by the Planckian ($\tau \sim \tau_\text{Pl}$) electrical transport observed in conventional metals such as copper above roughly the Debye temperature \cite{P2, Bruin804}. In this temperature regime the phonons are classical but the electrons are still degenerate. The electron-phonon scattering rate is $T$-linear due to the classical phonon cross section $A \sim \langle (\Delta x)^2 \rangle \propto k_B T$, from equipartition, while the quantum mechanical $\hbar$  originates purely in the Fermi velocity $v_F \sim t a/\hbar$, with $t$ the bandwidth and $a$ the lattice spacing. The faster the electrons, the more collisions with phonons occur per unit time.

Above roughly the Debye temperature in insulating crystals, the phonon umklapp scattering rate is again $T$-linear due to the classical phonon cross section. However, the relevant velocity is now the sound velocity $v_s$ which is also classical. The sound velocity is determined by quantities appearing in the atomic Hamiltonian --- atomic masses $M$, lattice spacing $a$ and spring constants $K$ --- without any explicit factors of $\hbar$. The corresponding heat transport is entirely classical from the phonon point of view.

We will show that there is, nonetheless, a quantum mechanical constraint on the sound velocity in a crystal.  The Heisenberg uncertainty principle together with the fact that atomic vibrations cannot make use of an energy scale greater than that holding the crystal together will lead to the bound $v_s \lesssim v_M \equiv (k_B T_M) a/\hbar$, with $T_M$ the crystal melting temperature.
The explicitly quantum mechanical `melting velocity' $v_M$ allows an analogy to the electron-phonon problem. We will see that this bound on the sound velocity implies the Planckian bound on the phonon umklapp scattering rate. In particular, Planckian scattering arises as the sound velocity gets closer to the melting velocity.
In Fig. \ref{fig:velocities} below we see that in several classes of compounds $v_M$ is a factor of roughly 5 to 19 times larger than $v_s$. This hierarchy between the velocities is microscopically grounded in the mass hierarchy $m/M \ll 1$, with $m$ the electron mass. The two velocities are somewhat correlated, but more important for our purposes is the spread in values of $v_M/v_s$. Fig. \ref{fig:ratios} shows that $\tau/\tau_\text{Pl} \approx \frac{1}{3} v_M/v_s$ for these compounds (with the exception of a class of highly conductive compounds with large $\tau$). In Fig. \ref{fig:ratios} we see that near-Planckian dynamics, due to small $v_M/v_s$, can occur for both complex and simple compounds.

In a nutshell: at high temperatures the phonon scattering rate $\tau^{-1} = v_s/\ell$, where the mean free path $\ell \propto 1/T$ is classical. Planckian phonon transport arises when the sound velocity $v_s$ is as large as is quantum mechanically possible.

\subsection*{Thermal transport above the Debye temperature}

A model Hamiltonian describing fluctuations of atoms about their equilibrium positions is
\be\label{eq:H}
H =  \sum_i \frac{p_i^2}{2 M} + \sum_{\langle ij \rangle} \left[\frac{K}{2} (x_i - x_j)^2 + \frac{\lambda}{6} (x_i - x_j)^3 + \cdots \right] \,.
\ee
In general there can be several atoms per unit cell, with different spring constants, masses and anharmonicities, and with $\langle ij \rangle$ corresponding to a sum over neighbours in the crystal structure.
Because $K, \lambda$ and the lattice spacing $a$ are all consistently determined from the same interatomic interactions, it is natural to expect that $K \sim \lambda a$. Indeed the measured Gr\"uneisen parameter is typically order one. According to the Lindemann criterion, the crystal will melt when fluctuations in the position of atoms extend to $\Delta x = c_L a$, with typically $c_L \approx 0.1 - 0.3$ (see e.g. \cite{simplemelt,elasticperov}). Therefore, while anharmonic couplings are order one in natural units, their contribution to physical processes below the melting temperature is suppressed by $\Delta x/a$. This allows us to keep only the leading anharmonic term in (\ref{eq:H}). We will, however, return to this point later.

The crystal will support both acoustic and optical phonon bands. The acoustic bands extend up to the Debye energy $k_B T_D = \hbar \omega_D \sim \hbar \sqrt{K/M}$.
Above the temperature $T_D$ the phonon states are macroscopically occupied and classical. The optical bands of simple crystals are at approximately this scale also and therefore become classical at roughly the same temperature. Let us be clear on the methodology here and throughout: our objective is not to reproduce numerical coefficients observed in particular materials. As mentioned above, this has already been achieved for both simple and complex materials. We wish to understand parametric constraints on transport observables in terms of quantities appearing in the atomic Hamiltonian.

While acoustic phonons typically carry most of the heat current, optical bands can play an important role in umklapp scattering. The anharmonic term in (\ref{eq:H}) allows for processes including a+a$\to$a, a+a$\to$o and a+o$\to$o. For temperatures $T \gtrsim T_D$, it is well known that the decay of acoustic phonons due to the these processes leads to a lifetime proportional to $T^{-1}$. In the Supplementary Material we give a quick derivation of this fact. The result, for a three dimensional crystal, can be written
\be
\frac{1}{\tau} \sim k_B T \frac{\lambda^2}{M K^2} \frac{Q^2}{v_s}a^3 \,. \label{eq:GammaMain}
\ee
Here $Q^2$ is the area of a surface in the Brillouin zone where phonon umklapp scattering is efficient, and $v_s$ is a `sound velocity' averaged over this surface. There are no $\hbar$s in (\ref{eq:GammaMain}). Given the Hamiltonian (\ref{eq:H}) it is an entirely classical result. Eq. (\ref{eq:GammaMain}) agrees with the expression in textbooks such as \cite{ziman}.

We can rewrite the result (\ref{eq:GammaMain}) in terms of the mean free path $\ell$ as
\be\label{eq:ell}
\frac{1}{\tau} = \frac{v_s}{\ell} \,, \qquad \ell \sim a^3 \frac{K}{\gamma^2 k_B T} \,.
\ee
We used the estimates $Q \sim 1/a$, $v_s^2 \sim a^2 K/M$ and introduced $\gamma \equiv \lambda\,a/{K}$. Here $\gamma$ is a dimensionless measure of the strength of the anharmonic interactions, and is roughly the high temperature Gr\"uneisen parameter. The estimate for $Q$ is rather crude and furthermore the nonlinearities contributing to the Gr\"uneisen parameter will not all contribute equally to umklapp processes. Nonetheless, the expression for $\ell$ in (\ref{eq:ell}) is physically transparent: $\ell = 1/(n A)$ with $n \sim 1/a^3$ the density of scatterers and the cross section $A \sim \gamma^2 \langle (\Delta x)^2 \rangle \sim \gamma^2 k_B T/K$. Here we are noting that $\gamma^2$ is the probability of interaction and equipartition requires $K \langle (\Delta x)^2 \rangle \sim k_B T$.

Moving beyond simple compounds, there will be an increasing number of optical bands available for the a+o$\,\rightarrow\,$o scattering process. In the Supplementary Material we show that 
these processes enhance the scattering rate by a factor of the number of accessible optical bands. The accessible optical bands, that can efficiently scatter acoustic phonons, are found to be those within roughly the energy range $\omega_D - 2 \omega_D$. There can be many such bands in complicated materials \cite{KA2}. This numerical factor will be folded into other numerical prefactors that we are not keeping track of, such as the portion of the Brillouin zone available for umklapp scattering and the difference between typical interatomic distances and the size of the unit cell. We will see that the only role of these numerical factors will be to distinguish a class of materials with anomalously long mean free paths compared to the rest (diamond, silicon, GaAs, BeO, etc. See Fig. \ref{fig:ratios}).

\subsection*{Saturation and the Slack-Kittel bound}

It will be instructive to differentiate the logic behind our Planckian bound from that of a distinct bound that has been conjectured for phonon transport.

The result (\ref{eq:ell}) for the decay rate leads to the thermal conductivity $\kappa = c D = c v_s^2 \tau \sim 1/T$ for $T \gtrsim T_D$. The specific heat $c$ is approximately constant at these temperatures. As the temperature is increased further still, two possible behaviors are observed experimentally. Firstly, that $\kappa \sim 1/T$ up to the melting temperature $T_M$. In other cases, $\kappa$ saturates to a constant value at a temperature $T_\text{sat} < T_M$ \cite{slack, PhysRevB.29.2884}.\footnote{There can also be upturns in $\kappa$ at high temperatures due to the onset of radiative heat transfer.}

Saturation is observed to occur when the mean free path $\ell$ approaches the interatomic spacing $a$. This is also of the order of the shortest phonon wavelength. A constant mean free path of this magnitude is characteristic of glasses and disordered solids \cite{PhysRev.75.972}. Furthermore, controlled disordering of crystals is found to interpolate between crystalline and glassy behavior \cite{PhysRevB.46.6131}.
Taken together, these facts are suggestive of a `Slack-Kittel' bound $\ell \gtrsim a$. 

From the expression (\ref{eq:ell}), the temperature at which $\ell \sim a$ is found to be $k_B T \sim \gamma^2 K a^2$. This is above the estimated melting temperature $k_B T_M \sim c_L^2 K a^2$. Therefore saturation can only be observed ($T_\text{sat} < T_M$) with favorable numerical coefficients that can overcome the factors of $c_L$. Recall from below (\ref{eq:H}) that the factors of $c_L$ are also responsible for suppressing higher order anharmonic terms below the melting temperature: the crystal melts before atomic spatial fluctuations become large. It is plausible, then, that transport with $\ell \sim a$ in simple insulators is strongly anharmonic and formally beyond the Peierls-Boltzmann framework \cite{PhysRevB.46.6131,PhysRevB.49.9073}. Because the Slack-Kittel bound and saturation occur within a purely classical phonon transport regime, they can be probed by numerical simulation of classical atoms \cite{PhysRevB.34.5058}. Such simulations have seen conductivity saturation at high temperatures \cite{PhysRevB.82.224305}, associated to phonon anharmonicity.

The Planckian bound that we will now discuss is orthogonal to the Slack-Kittel bound in the following precise sense. Write the scattering rate as $1/\tau = v_s/\ell$. The Slack-Kittel bound is the statement that $\ell \gtrsim \ell^\text{min}$. The Planckian bound will instead come from the statement that $v_s \lesssim v_s^\text{max}$. Our discussion of the Planckian bound will focus on the regime $T_D \lesssim T \lesssim \text{min}(T_\text{sat},T_M)$, where the mean free path $\ell \sim 1/T$ is given by (\ref{eq:ell}). While the Slack-Kittel bound on $\ell$ bounds the magnitude of the thermal diffusivity $D$, the bound on the velocity bounds the slope of $D^{-1} \sim T$.

\subsection*{The melting velocity}

We now describe a quantum mechanical upper bound on the sound velocity. In quantum mechanical systems with a finite dimensional on-site Hilbert space (e.g. spins or fermions) and bounded local interactions on a lattice with spacing $a$ there is a maximal Lieb-Robinson velocity $v_\text{LR}$ that bounds all physical velocities $v$ \cite{Lieb1972}:
\be\label{eq:LR}
v \leq v_\text{LR} \sim \frac{J a}{\hbar} \,.
\ee
Here $J$ is the maximal coupling between neighbouring sites on the lattice. For example, for free fermions $J \sim t$, the bandwidth, and the Lieb-Robinson velocity is roughly the Fermi velocity $v_F$ \cite{Hartman:2017hhp}. As we recalled above, the inverse $\hbar$ in the Fermi velocity is indeed responsible for the Planckian transport times observed due to electron-phonon scattering above the Debye scale in conventional metals such as copper, with $v_s \to v_F$ in the first equation in (\ref{eq:ell}).

The atomic Hamiltonian (\ref{eq:H}) does not fall under the auspices of the Lieb-Robinson theorem because the full single-particle Hilbert space is not bounded.\footnote{Extensions of the theorem to oscillator lattice systems, such as \cite{Nachtergaele2009}, bound information transfer due to incoherent hopping between sites, but do not constrain the velocity of processes such as sound waves that involve the motion of an atom at a given site in an essential way.} However, a Lieb-Robinson-like bound on the sound speed is obtained as follows. In the crystalline state, the total energy of any given atom should not exceed the melting temperature
\bea\label{eq:boundM}
k_B T_M \gtrsim \frac{p^2}{2 M} + \frac{K}{2} (x - x_\text{eq})^2 \,,
\eea
otherwise the atom would no longer be bound to the crystal (according to the Lindemann criterion). Here $x_\text{eq}$ is the classical equilibrium position of the atom. We can simplify (\ref{eq:boundM}) by writing
\be
\frac{p^2}{2 M} + \frac{K}{2} (x - x_\text{eq})^2  \, \geq \, \frac{1}{2 M} (\Delta p)^2 + \frac{K}{2} [\Delta (x - x_\text{eq})]^2 \, \geq \,  \sqrt{\frac{K}{M}} \Delta p \cdot \Delta (x - x_\text{eq}) \,. \label{eq:two}
\ee
The first step used the definition of the variance. The second step follows from $a^2 + b^2 = (a-b)^2 + 2 ab \geq 2ab$. Now, $p$ is conjugate to $x - x_\text{eq}$ because $x_\text{eq}$ only depends on the positions of other atoms. Therefore, putting (\ref{eq:boundM}) and (\ref{eq:two}) together and using the Heisenberg uncertainty principle gives
$k_B T_M \gtrsim \hbar \sqrt{K/M} \sim k_B T_D$. The mass cannot become too small, else the crystal would spontaneously melt due to quantum mechanical zero point motion of the atoms. Multiplying by a characteristic interatomic spacing $a$, this inequality is equivalent to a bound on the sound speed
\be\label{eq:vM}
v_s \lesssim v_M \equiv \frac{(k_B T_M) a}{\hbar} \,.
\ee
Here we introduced the `melting velocity' $v_M$. This bound is formally analogous to the Lieb-Robinson bound (\ref{eq:LR}), with the melting temperature playing the role of the energy scale $J$. This suggests the intuitive picture that the largest energy scale available for the motion of phonons is that responsible for holding the crystal together. 

The ratio of sound and melting velocities can be estimated as follows. The sound velocity is set by the atomic mass $M$ while the energetics of melting microscopically depends on the much smaller electron mass $m$. One roughly expects
\be\label{eq:vrat}
\frac{v_s}{v_M} \sim \frac{\hbar}{c_L^2 K a^3} \frac{a K^{1/2}}{M^{1/2}} \sim \frac{\hbar}{c_L^2 a^2 (KM)^{1/2}} \sim \frac{1}{c_L^2} \left(\frac{m}{M} \right)^{1/2} \,.
\ee
In the final step we estimated $K a^2 \sim \hbar^2/(m a^2)$ by equating the binding energy with the kinetic energy of the electrons. The same estimate is obtained in an ionic crystal with $K a^2 \sim e^2/a$, because $a$ is of the order of the Bohr radius. The large numerical factor of $1/c_L^2$ in (\ref{eq:vrat}) opposes the mass hierarchy $m \ll M$.
Such numerical factors are necessary in order for the 
velocity bound (\ref{eq:vM}) to come close to saturation.

\begin{figure}[h!]
    \centering
    \includegraphics[width=\textwidth]{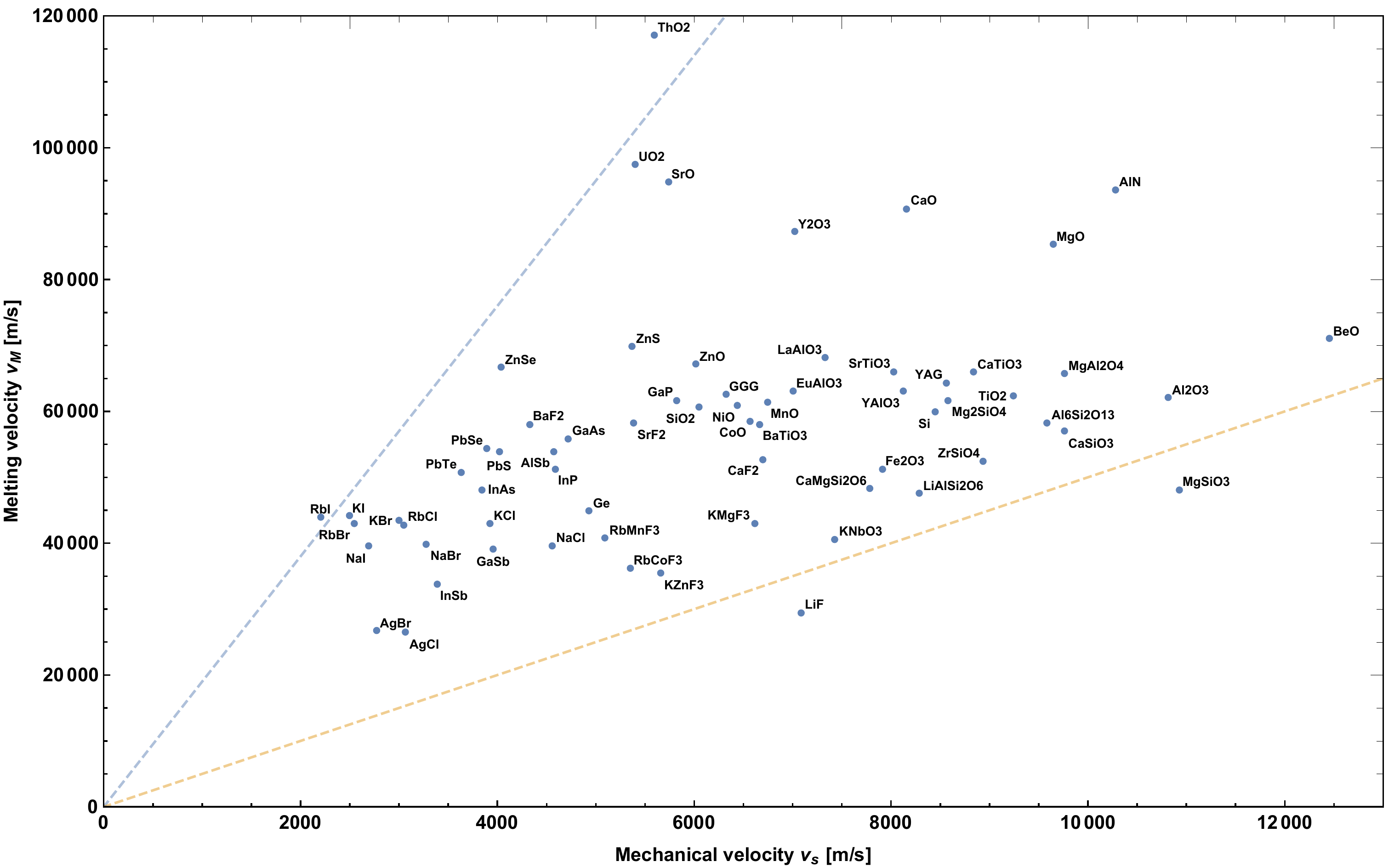}
    \caption{Melting velocity versus sound velocity for various classes of non-metallic compound. The precise definitions of the velocities are given in the main text. The dashed lines show $v_M = 19 v_s$ (upper) and $v_M = 5 v_s$ (lower).}
    \label{fig:velocities}
\end{figure}

A plot of $v_s$ versus $v_M$ for several families of insulating crystals is shown in Fig. \ref{fig:velocities}. The most important information in this figure is that for these compounds the ratio $v_M/v_s$ runs from about 5 to about 19. These are the dashed lines shown in the figure. The values plotted are tabulated in the Supplementary Material along with references. The sound velocity has been computed from measured values of the bulk modulus $K$, shear modulus $G$ and density $\rho$ according to $v_s^2 = (K + \frac{4}{3} G)/\rho$. This is the velocity of longitudinal sound waves in an isotropic solid, and defines a characteristic `mechanical' velocity more generally. The melting velocity has been computed from the observed melting temperature $T_M$, with the length $a$ taken to be the average interatomic distance. The latter is obtained from the observed density $\rho$ and the molar mass. The quantities $K,G,\rho$ are evaluated at room temperature and atmospheric pressure, as this is where the most data is available.

The compounds in Fig.\ref{fig:velocities} include alkali halides, oxides with varying degree of complexity and several types of semiconductor. We have focused on materials for which we have been able to find high temperature thermal transport data, as we will shortly correlate high temperature transport behavior with the ratio $v_M/v_s$.

\subsection*{From the velocity bound to the Planckian bound}

If we use the velocity bound (\ref{eq:vM}) in the scattering rate (\ref{eq:ell}) and furthermore drop all dimensionless numerical factors (including $\gamma^2$, $c_L^2$ and phase space factors in the scattering computation, the correctness of this procedure will be verified a posteriori), then we obtain a Planckian bound on the phonon lifetime
\be\label{eq:tauP}
\frac{\tau}{\tau_\text{Pl}} \sim \frac{v_M}{v_s} \gtrsim 1 \,.
\ee
That is, the Lieb-Robinson type bound (\ref{eq:vM}) on the sound velocity implies a Planckian bound (\ref{eq:planck}) on scattering. To our knowledge this is the first Planckian bound that has been theoretically grounded.

\begin{figure}[h!]
    \centering
    \includegraphics[width=\textwidth]{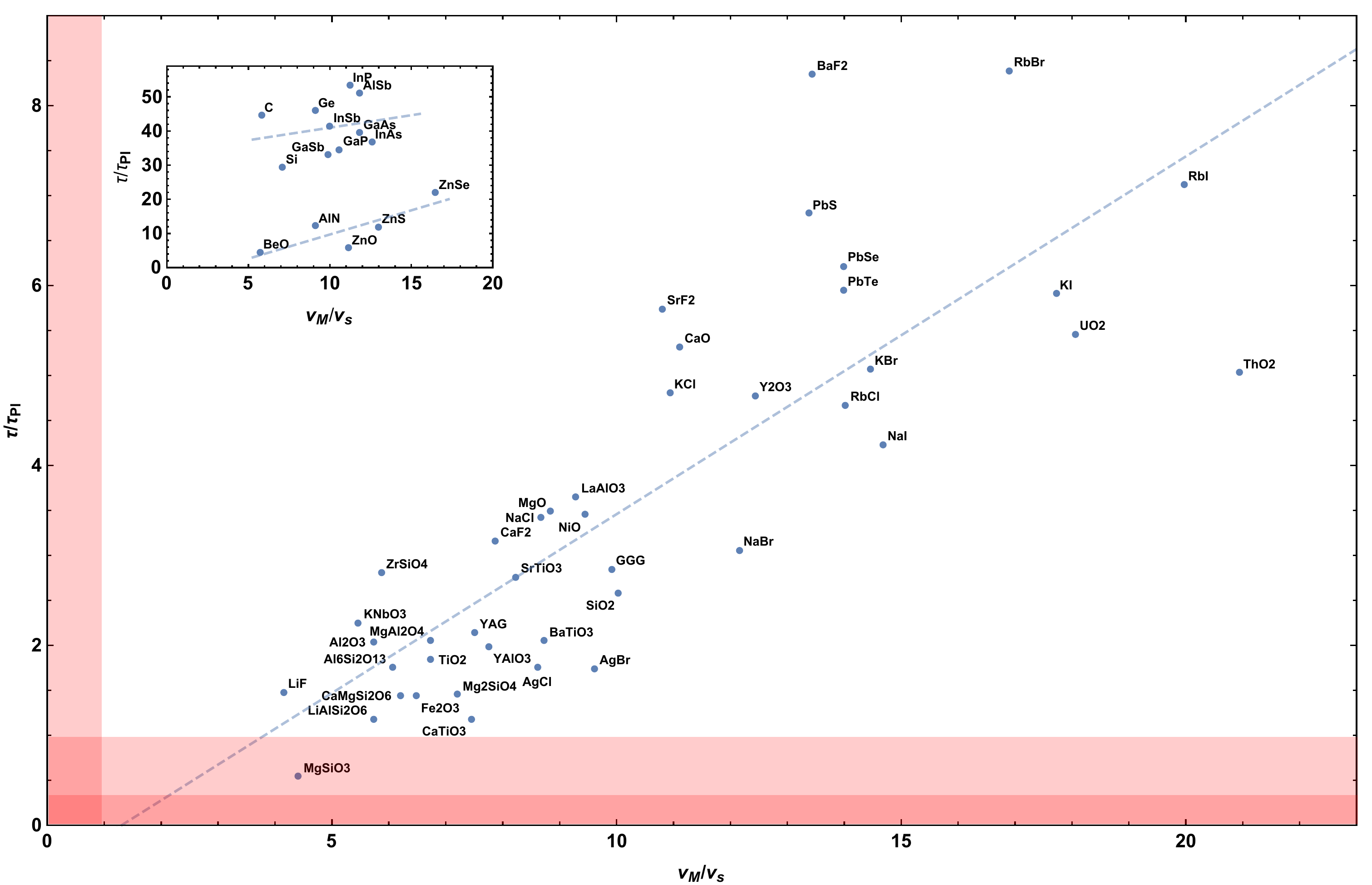}
    \caption{Ratio of timescales $\tau/\tau_\text{Pl}$ versus the ratio of velocities $v_M/v_s$. The inset shows `adamantine' crystals with a large mean free path, discussed in the main text. Linear fits are shown as guides to the eye. The velocity and Planckian bounds are shown as shaded regions. We have also shown the Planckian bound for $\tau' = 3 \tau$, see main text.}
    \label{fig:ratios}
\end{figure}

The most basic assertion of (\ref{eq:tauP}) is that in the high temperature regime of phonon umklapp scattering, the ratio of velocities $v_M/v_s$ should determine the scattering ratio $\tau/\tau_\text{Pl}$.
The closer the sound velocity to the melting velocity, the closer the scattering rate to the Planckian bound. Fig. \ref{fig:ratios} shows $\tau/\tau_\text{Pl}$ versus $v_M/v_s$ for several families of non-metallic compounds, revealing precisely this trend at work.

In Fig. \ref{fig:ratios} the values of $v_M/v_s$ are taken from Fig. \ref{fig:velocities} while the lifetime $\tau$ is obtained from the thermal diffusivity as $D = v_s^2 \tau$. Where direct measurements of thermal diffusivity are available, these have been used. Otherwise, diffusivity has been obtained from the measured thermal conductivity and specific heat. Data are tabulated in the Supplementary Material, along with references. The diffusivities have been evaluated at high temperatures, where experimentally $D \sim 1/T$ is seen to either hold exactly or to be a good approximation. The extracted ratio $\tau/\tau_\text{Pl}$ does not have a significant dependence on temperature in such regimes. Recall that $v_s$ is the room temperature sound velocity. Use of the velocity at the same high temperature at which transport is measured would be logically more satisfying, but is not expected to introduce a strong temperature dependence. In this regard the methodology behind Fig. \ref{fig:ratios} is similar to that in \cite{Bruin804}.

\subsection*{Discussion}

Fig. \ref{fig:ratios} shows that --- with the exception of the highly conductive `adamantine' compounds to be discussed shortly --- the sound velocity of materials determines whether Planckian thermal transport will arise.
The sound velocity should be measured relative to the melting velocity.
While the Planckian (low $\tau/\tau_\text{Pl}$) end of the plot is mostly populated by somewhat complex oxides, there are also simple materials such as LiF that appear. The logic leading to the velocity bound (\ref{eq:vM}) suggests that LiF should be somewhat close to spontaneous melting. Indeed, measured Debye-Waller factors show that the mean square atomic vibrational amplitude has a weaker temperature dependence in LiF than in other alkali halides, such that the amplitude of zero temperature quantum vibrations is a larger fraction of the amplitude that melts the crystal \cite{Martin_1977}.

Zero point motion is significant in LiF because its constituent atoms are light. Correspondingly, LiF has a large sound velocity.
Note that a large sound velocity favors Planckian scattering, even while making the thermal conductivity $\kappa \sim c v_s^2 \tau$ large. Thus, for example, the two alkali halides LiF and RbI appear at opposite ends of Fig. \ref{fig:ratios}, despite having comparable thermal conductivities at high temperatures \cite{hof-simple}. The difference in sound velocities between these two materials is of the order predicted by the heavier mass of the constituent atoms of RbI.

While the linear fits in Fig. \ref{fig:ratios} should not be overinterpreted, given the finite number of materials considered, it is interesting that the main fit has a slope $\tau/\tau_\text{Pl} \approx \frac{1}{3} v_M/v_s$. This factor of $\frac{1}{3}$ would have been absent if we had defined a timescale $\tau'$ via $D = \frac{1}{3} v_s^2 \tau'$ (as opposed to $D = v_s^2 \tau$). That is, this factor of $\frac{1}{3}$ is natural in three dimensions and suggests that the melting velocity indeed controls the proximity to Planckian scattering, without any additional large numerical factors. Our systematic neglect of numerical factors in the scattering rate is thus seen to be justified, they tend to cancel out on average. The mean free path $\ell' = v_s \tau'$ is then found to be
\be\label{eq:lprime}
\ell' \approx \frac{T_M}{T} a \,.
\ee
Recall that $a$ has been defined as the average interatomic spacing. Eq. (\ref{eq:lprime}) is consistent with the observation that mean free paths typically approach the interatomic spacing close to the melting temperature \cite{slack}.

The inset of Fig. \ref{fig:ratios} shows a class of crystals for which $\ell'$ is significantly greater than the trend (\ref{eq:lprime}) obeyed by most of the compounds. These crystals appear to cluster into two groups in the figure, but we are not aware of an explanation for this fact. These `adamantine' materials all have zincblende or wurtzite crystal structures and several of them are well-known to have anomalously high thermal conductivities \cite{slack, slack2}. It may
be interesting to revisit the properties of this class of crystals from the point of view of their anomalously large hierarchy between $\tau$ and $\tau_\text{Pl}$. Near Planckian rocksalt compounds such as   LiF or AgBr in Fig. \ref{fig:ratios} show that this phenomenon (of relatively large mean free paths) cannot be entirely due to crystallographic simplicity.

\subsection*{Acknowledgements}

We have benefited greatly from insightful comments and criticism from Kamran Behnia, Aharon Kapitulnik, Steve Kivelson and Jan Zaanen. We also thank Zhaoyu Han and Darius Shi for discussions on related topics. This work is supported by the Department of Energy, Office of Basic Energy Sciences, under Contract No. DEAC02-76SF00515. CHM is supported by an NSF graduate fellowship.

\providecommand{\href}[2]{#2}\begingroup\raggedright\endgroup

\newpage

\appendix

\noindent {\bf \LARGE Supplementary Material}

\section{Scattering above the Debye temperature}

\subsection*{Simple compounds}

It will be convenient to work with the Lagrangian for the normal phonon modes.
If $a_{sq}^\dagger$ creates a phonon with wavevector $q$ in the band $s$, then letting $b_{sq} \equiv a_{sq} + a_{s-q}^\dagger$ one
has, from the Hamiltonian (\ref{eq:H}),
\be\label{eq:L}
L = \frac{\hbar}{2} 
\sum_{sq} \frac{\omega^2 - \omega_{sq}^2}{2 \omega_{sq}} b_{sq} b_{s-q} + 
\frac{\lambda \, \hbar^{3/2}}{6 (K M)^{3/4}} \sqrt{\frac{a^{3}}{V}} \sum_{s_i,q_i} f^{s_1s_2s_3}_{q_1 q_2 q_3} b_{s_1q_1} b_{s_2q_2} b_{s_3q_3} \,.
\ee
Here $a$ is the lattice spacing and $V$ the total volume. The precise form of the modes $\omega_{sq}$ and the dimensionless function $f^{s_1s_2s_3}_{q_1 q_2 q_3}$ depend on the lattice structure. Scatterings are only allowed if they conserve crystal momentum up to a reciprocal lattice vector.

From the Lagrangian (\ref{eq:L}) the retarded phonon Green's function is
\be
D^R(\omega,q) = \frac{2 \omega_q}{\omega^2 - \omega_q^2} = \frac{1}{\omega - \omega_q} - \frac{1}{\omega + \omega_q} \,.
\ee
While the kinematics of the scattering responsible for a finite thermal conductivity requires two phonon bands, for simple crystals the parameters will be similar for the different bands and so we will not keep track of the band label. The phonon self-energy is then given to lowest order as
\bea\label{eq:ImSig}
\lefteqn{\text{Im}\,\Sigma(\omega,k) =} \\
&& - \frac{\lambda^2 \hbar}{36 (K M)^{3/2}} \frac{a^3}{V} \sum_q f^2_{k,q,k+q} \int_{-\infty}^\infty \frac{d\Omega}{\pi} \text{Im}\, D^R(\Omega,q) \text{Im} D^R(\omega+\Omega,k+q) \frac{n_B(\omega+\Omega) n_B(-\Omega)}{n_B(\omega)} \,. \nonumber
\eea
Here $n_B$ is the Bose-Einstein distribution. At temperatures $T \gtrsim T_D$ all of these factors are in the high temperature limit, so that $n_B(\omega) \approx k_B T/(\hbar \omega)$. Thus, doing the $\Omega$ integral
\be\label{eq:ImSig2}
\text{Im}\,\Sigma(\omega,k) = \frac{\pi k_B T \lambda^2}{36 (K M)^{3/2}} \frac{a^3}{V}\sum_{q}  \frac{\omega f^2_{k,q,k+q} }{\omega_{k+q}\omega_q} \Big(
\delta(\omega + \omega_q - \omega_{k+q}) + \delta(\omega - \omega_q - \omega_{k-q}) \Big) \,.
\ee
Note that there are no $\hbar$'s remaining in this expression. The high temperature regime is classical.

In a three dimensional crystal, then
\be
\frac{1}{\tau} = \text{Im}\,\Sigma \sim k_B T \frac{\lambda^2}{M K^2} \frac{Q^2}{v_s}a^3  \,. \label{eq:Gamma}
\ee
In going from (\ref{eq:ImSig2}) to (\ref{eq:Gamma}) we have set $\frac{1}{V} \sum_q \delta(\w \pm \omega_q - \omega_{k\pm q}) \to Q^2/v_s$ in three dimensions, so that $Q^2$ is the area of a surface in the Brillouin zone where phonon umklapp scattering is efficient, and $v_s$ is a `sound velocity' averaged over this surface. In this average we furthermore used a typical frequency $\omega \sim \sqrt{K/M}$. Eq. (\ref{eq:Gamma}) is Eq. (\ref{eq:GammaMain}) in the main text.

\subsection*{Complex compounds}

In more complex crystals, acoustic phonons will typically still dominate the heat current but now the presence of a large number of optical bands means that the process a+o$\,\rightarrow\,$o makes available a large scattering phase space for the acoustic phonons.
Restoring the band dependence of the coupling $f$ and of the dispersions, (\ref{eq:ImSig2}) becomes
\bea
\lefteqn{\text{Im}\,\Sigma_{a}(\omega,k) =} \\
& & - \frac{\lambda^2 \hbar\,\pi}{36 (K M)^{3/2}} \frac{a^3}{V} \sum_{b_1,b_2}\sum_q |f^{\,a,b_1,b_2}_{k,q,k+q}|^2 
\delta(\omega+\omega_q^{b_1}-\omega_{k+q}^{b_2}) 
\frac{n_B(\omega+\omega_q^{b_1}) n_B(-\omega_q^{b_1})}{n_B(\omega)} \,. \nonumber
\eea
Here $a$ denotes an acoustic band and $b_1,b_2$ are optical bands. Ratios of differing atomic masses and spring constants are all subsumed into the $f$ couplings; $M$ and $K$ are typical magnitudes of these quantities that set the overall scale. Only the $\delta$-functions corresponding to a+o$\,\rightarrow\,$o processes are retained (a\,$\to$\,o+o is not possible on shell). We have not yet expanded the Bose-Einstein factors because the temperatures of interest, while greater than $T_D$, can be in the middle of the plethora of optical phonon bands, and not all bands will be classical.

We will see now that the presence of non-classical optical bands at high energies (potentially greater than the temperatures probed) does not spoil the $T$-linear scattering rate. Only bands that are sufficiently close in energy to the acoustic bands are able to scatter the acoustic phonons efficiently. This is because the occupancy of the high energy bands is suppressed by Bose-Einstein factors relative to the acoustic bands. A series of reasonable approximations brings out the essential physics. Firstly, the optical bands have small bandwidths and can be approximated as Einstein phonons at the average band frequency, so that $n_B(\omega_q^{b}) \to n_B(\braket{\omega_{q}^{b}})$. Secondly, kinematic constraints mean that given $a$ and $b_1$ fixed, only an order one number of $b_2$ bands are accessible for an $a + b_1 \to b_2$ process. This effectively means that there is only one sum over bands. Thirdly, this single sum over a large number of bands can be approximated by an integral: $\sum_{b_1}
\rightarrow \frac{1}{\Delta \omega}\int_{\omega^{\text{o}}_{\text{min}}}^{\omega^{\text{o}}_{\text{max}}}d\Omega$. Here $\Delta \omega$ is the average separation between optical bands and $\omega^{\text{o}}_{\text{min/max}}$ are the minimum/maximum frequencies of optical bands. Thus we obtain
\bea
\text{Im}\,\Sigma_{a}(\omega,k) \sim
- \frac{\lambda^2 \hbar}{(K M)^{3/2}} \frac{Q^2 a^3}{v_\text{o}} \frac{1}{\Delta\omega}
\int_{\omega^{\text{o}}_{\text{min}}}^{\omega^{\text{o}}_{\text{max}}}d\Omega
\,\frac{n_B(\omega + \Omega)n_B(-\Omega)}
{n_B(\omega)}
\,, \label{eq:nnn}
\eea
where we again let $\frac{1}{V}\sum_q \delta(\omega) \sim Q^2/v_\text{o}$. While $Q^2$ is again a surface of the Brillouin zone where umklapp scattering is allowed, $v_\text{o}$ is now a typical optical phonon velocity.

In (\ref{eq:nnn}) we have, for the complex materials with many optical bands, $\omega^\text{o}_\text{max} \gg \omega^\text{o}_\text{min} \sim \omega_D \sim \omega$. The temperature $T$ is greater than the scales set by $\omega^\text{o}_\text{min}, \omega_D$ and $\omega$ but could be greater or smaller than $\omega^\text{o}_\text{max}$. The integral in (\ref{eq:nnn}) can be done exactly, and in this parameter regime goes like $k_B T/\hbar \times \log [(\omega^\text{o}_\text{min}+\omega)/\omega^\text{o}_\text{min}] \sim k_B T/\hbar$. This happens because, as anticipated above, the acoustic phonons are only efficiently scattered by optical phonons with frequencies close to $\omega \sim \omega_D$. This occurs because the Bose-Einstein factors in the integrand in (\ref{eq:nnn}) are all dominated by the small frequency regime wherein $n_B(x) \sim 1/x$. Therefore we obtain
\bea\label{eq:SigCom}
\text{Im}\,\Sigma_{a} \sim 
\frac{\lambda^2}{(K M)^{3/2}} \frac{Q^2 a^3}{v_\text{o}} \frac{k_BT}{\Delta\omega} \sim \frac{\omega_D}{\Delta \omega} \frac{v_s}{v_\text{o}} \, \text{Im} \, \Sigma\,.
\eea
The final expression shows that the phase space due to a+o$\,\rightarrow\,$o processes in complex materials has increased the decay rate relative to the result (\ref{eq:Gamma}) in simple crystals. The increased scattering rate is qualitatively consistent with previous estimates of the effect of crystal complexity on transport, e.g. \cite{slack}. For distorted perovskites there are about 10 bands between $\omega_D$ and $2 \omega_D$, so that $\omega_D/\Delta \omega \sim 10$ \cite{phonon2, doi:10.1143/JPSJ.72.1418}. The optical velocity $v_o$ will furthermore be some fraction of the acoustic velocity $v_a$.

\section{Material data}
\label{sec:data}

The table below has been used to make the figures in the main text. The quantities appearing in the table are: bulk modulus $K$, shear modulus $G$, density $\rho$, molar mass $M$, average interatomic distance $\overline a$ (obtained from the density, molar mass and number of atoms per molecule), melting temperature $T_M$, sound velocity $v_s$ computed as described in main text, melting velocity $v_M$ computed as described in main text, diffusivity $D$, temperature $T$ at which diffusivity is evaluated and transport time $\tau$ computed as described in the main text. Where direct diffusivity data is available, this has been used. Otherwise the diffusivity has been computed from the thermal conductivity and specific heat. The diffusivity is absent for a few compounds for which we did not find transport measurements in a high temperature regime with at least an approximate $1/T$ dependence. References for the elastic moduli, density, melting temperature and diffusivity are given below the table.

\small
\begin{longtable}{l|llllllllllll}
  & \makecell{$K$ \\ {\scriptsize GPa}} & \makecell{$G$ \\ {\scriptsize GPa}} & \makecell{ $\rho$ \\ \scriptsize  g/cm$^3$} & \makecell{$M$ \\ \scriptsize g/mol} & \makecell{$\overline a$ \\ \scriptsize  A} & \makecell{$T_M$\\ \scriptsize  K} & \makecell{$v_s$\\ \scriptsize  m/s} & \makecell{$v_M$ \\ \scriptsize  m/s} & $\displaystyle \frac{v_M}{v_s}$ & \makecell{$D$\\ \scriptsize  mm$^2$/s} & \makecell{$T$\\ \scriptsize  K} & $\displaystyle \frac{\tau}{\tau_\text{Pl}}$  \\
  \hline
Si        & 98              & 51               & 2.33            & 28.1               & 2.716                            & 1687   & 8441              & 59925                  & 7.10    & 20     & 800   & 29.4       \\
Ge        & 75              & 41               & 5.32            & 72.7               & 2.831                            & 1211   & 4937              & 44847                  & 9.08    & 11.4     & 750   & 45.8       \\
Diamond   & 443             & 536              & 3.512           & 12                 & 1.784                            & 4500   & 18156             & 104987                 & 5.78    & 140    & 800   & 44.5       \\
GaP       & 88              & 39               & 4.13            & 101                & 2.728                            & 1730   & 5822              & 61736                  & 10.60   & 17     & 530   & 34.7       \\
GaAs      & 76              & 32               & 5.32            & 145                & 2.829                            & 1511   & 4723              & 55905                  & 11.84   & 11.2     & 600   & 39.5       \\
GaSb      & 56              & 24               & 5.62            & 191                & 3.044                            & 985    & 3957              & 39226                  & 9.91    & 4.94      & 800   & 33       \\
InAs      & 58              & 19               & 5.66            & 190                & 3.032                            & 1215   & 3837              & 48186                  & 12.56   & 5.91      & 700   & 36.7       \\
InP       & 71              & 22               & 4.78            & 146                & 2.938                            & 1335   & 4582              & 51305                  & 11.20   & 12.2     & 700   & 53.3       \\
InSb      & 46              & 15               & 5.77            & 237                & 3.243                            & 798    & 3382              & 33850                  & 10.01   & 6.59      & 550   & 41.4       \\
AlSb      & 58              & 22.5             & 4.218           & 149                & 3.084                            & 1338   & 4568              & 53974                  & 11.82   & 10.2     & 800   & 51.1       \\
AlN       & 198             & 110              & 3.26            & 41                 & 2.186                            & 3273   & 10282             & 93577                  & 9.10    & 5.36      & 1823  & 12.1       \\
ZnS       & 74              & 33               & 4.084           & 97.5               & 2.706                            & 1973   & 5375              & 69845                  & 12.99   & 4.10      & 643   & 11.9       \\
ZnSe      & 62              & 18               & 5.264           & 144                & 2.832                            & 1798   & 4042              & 66605                  & 16.48   & 6.84      & 400   & 21.9       \\
ZnO       & 143             & 47               & 5.675           & 81.4               & 2.284                            & 2247   & 6020              & 67121                  & 11.15   & 1.46      & 1073  & 5.67        \\
BeO       & 251             & 162              & 3.01            & 25                 & 1.903                            & 2850   & 12456             & 70959                  & 5.70    & 3.42      & 1565  & 4.51        \\
NaBr      & 19.3            & 11.4             & 3.2             & 103                & 2.990                            & 1020   & 3283              & 39890                  & 12.15   & 0.28      & 900   & 3.06        \\
LiF       & 67.2            & 49.1             & 2.64            & 25.9               & 2.012                            & 1121   & 7089              & 29504                  & 4.16    & 0.63      & 900   & 1.48        \\
KCl       & 18.1            & 9.4              & 1.987           & 74.5               & 3.146                            & 1044   & 3926              & 42959                  & 10.94   & 0.63      & 900   & 4.81        \\
KBr       & 14.3            & 7.9              & 2.75            & 119                & 3.300                            & 1007   & 3005              & 43465                  & 14.46   & 0.50      & 700   & 5.07        \\
RbI       & 10.5            & 5.03             & 3.55            & 212                & 3.674                            & 915    & 2202              & 43970                  & 19.97   & 0.33      & 800   & 7.12        \\
RbBr      & 13              & 6.54             & 3.35            & 165                & 3.445                            & 955    & 2546              & 43038                  & 16.90   & 0.52      & 800   & 8.39        \\
RbCl      & 15.4            & 7.75             & 2.76            & 121                & 3.314                            & 988    & 3053              & 42830                  & 14.03   & 0.37      & 900   & 4.67        \\
SrO       & 87              & 58               & 5               & 104                & 2.585                            & 2804   & 5733              & 94806                  & 16.54   &           &       &             \\
KI        & 11.3            & 6.03             & 3.12            & 166                & 3.535                            & 954    & 2490              & 44113                  & 17.72   & 0.40      & 700   & 5.91        \\
AgCl      & 41.6            & 8.10             & 5.56            & 143                & 2.774                            & 728    & 3070              & 26419                  & 8.61    & 0.21      & 600   & 1.75        \\
AgBr      & 37.8            & 8.89             & 6.47            & 188                & 2.890                            & 705    & 2770              & 26647                  & 9.62    & 0.17      & 600   & 1.74        \\
NaCl      & 25              & 15               & 2.163           & 58.4               & 2.820                            & 1074   & 4561              & 39612                  & 8.68    & 0.68      & 800   & 3.42        \\
NaI       & 15.1            & 8.6              & 3.67            & 150                & 3.238                            & 933    & 2691              & 39512                  & 14.69   & 0.39      & 600   & 4.23        \\
PbS       & 63              & 45               & 7.61            & 239                & 2.965                            & 1386   & 4020              & 53761                  & 13.37   & 1.40      & 600   & 6.80        \\
PbSe      & 54              & 52               & 8.15            & 286                & 3.077                            & 1351   & 3890              & 54379                  & 13.98   & 1.20      & 600   & 6.22        \\
PbTe      & 41              & 50               & 8.16            & 335                & 3.242                            & 1197   & 3632              & 50767                  & 13.98   & 1.00      & 600   & 5.95        \\
MgO       & 160             & 130              & 3.584           & 40.3               & 2.106                            & 3098   & 9644              & 85329                  & 8.85    & 1.71      & 1452  & 3.49        \\
CaO       & 115             & 81               & 3.346           & 56.1               & 2.406                            & 2886   & 8164              & 90812                  & 11.12   & 2.53      & 1073  & 5.32        \\
NiO       & 205             & 59               & 6.828           & 74.7               & 2.087                            & 2230   & 6446              & 60863                  & 9.44    & 0.86      & 1273  & 3.45        \\
MnO       & 153             & 68               & 5.365           & 70.9               & 2.222                            & 2115   & 6739              & 61476                  & 9.12    &           &       &             \\
CoO       & 183             & 71               & 6.438           & 74.9               & 2.130                            & 2103   & 6567              & 58585                  & 8.92    &           &       &             \\
ThO$_2$      & 178             & 94               & 9.7             & 264                & 2.470                            & 3623   & 5592              & 117045                 & 20.93   & 0.82      & 1473  & 5.03        \\
TiO$_2$      & 215             & 112              & 4.26            & 79.9               & 2.182                            & 2185   & 9248              & 62350                  & 6.74    & 0.82      & 1473  & 1.84        \\
UO$_2$       & 209             & 83               & 10.97           & 270                & 2.388                            & 3120   & 5398              & 97472                  & 18.06   & 0.96      & 1273  & 5.46        \\
SiO$_2$      & 38              & 44               & 2.648           & 60.1               & 2.325                            & 1995   & 6042              & 60664                  & 10.04   & 1.20      & 600   & 2.58        \\
CaF$_2$      & 86.3            & 42.4             & 3.181           & 78.1               & 2.386                            & 1691   & 6701              & 52785                  & 7.88    & 3.61      & 300   & 3.16        \\
SrF$_2$      & 71              & 39               & 4.24            & 126                & 2.543                            & 1750   & 5386              & 58216                  & 10.81   & 4.25      & 300   & 5.74        \\
BaF$_2$      & 58              & 25               & 4.886           & 175                & 2.706                            & 1641   & 4324              & 58096                  & 13.44   & 3.98      & 300   & 8.35        \\
Fe$_2$O$_3$     & 207             & 91               & 5.254           & 160                & 2.163                            & 1812   & 7905              & 51258                  & 6.48    & 0.65      & 1056  & 1.44        \\
Al$_2$O$_3$     & 252             & 162              & 3.999           & 102                & 2.039                            & 2326   & 10818             & 62022                  & 5.73    & 1.03      & 1771  & 2.04        \\
Y$_2$O$_3$      & 152             & 72.6             & 5.05            & 226                & 2.459                            & 2712   & 7019              & 87220                  & 12.43   & 6.00      & 300   & 4.78        \\
CaSiO$_3$    & 227             & 125              & 4.13            & 172                & 2.400                            & 1813   & 9763              & 56926                  & 5.83    &           &       &             \\
SrTiO$_3$    & 174             & 117              & 5.12            & 183                & 2.281                            & 2213   & 8028              & 66034                  & 8.23    & 1.70      & 800   & 2.76        \\
YAlO$_3$     & 204             & 112              & 5.35            & 164                & 2.167                            & 2223   & 8127              & 63022                  & 7.75    & 0.59      & 1700  & 1.99        \\
BaTiO$_3$    & 177             & 68               & 6.04            & 233                & 2.340                            & 1898   & 6657              & 58093                  & 8.73    & 0.77      & 900   & 2.05        \\
KMgF$_3$     & 75              & 47               & 3.15            & 120                & 2.330                            & 1413   & 6611              & 43068                  & 6.51    &           &       &             \\
KZnF$_3$     & 78              & 38               & 4.02            & 161                & 2.369                            & 1143   & 5657              & 35424                  & 6.26    &           &       &             \\
RbMnF$_3$    & 67              & 34               & 4.32            & 197                & 2.474                            & 1259   & 5099              & 40745                  & 7.99    &           &       &             \\
RbCoF$_3$    & 80              & 42               & 4.76            & 201                & 2.412                            & 1148   & 5345              & 36213                  & 6.77    &           &       &             \\
CaTiO$_3$    & 177             & 104              & 4.04            & 136                & 2.236                            & 2253   & 8839              & 65898                  & 7.45    & 0.78      & 900   & 1.18        \\
EuAlO$_3$    & 203             & 114              & 7.25            & 226                & 2.179                            & 2213   & 6998              & 63090                  & 9.02    &           &       &             \\
LaAlO$_3$    & 195             & 117              & 6.52            & 214                & 2.217                            & 2350   & 7337              & 68157                  & 9.29    & 0.94      & 1600  & 3.65        \\
KNbO$_3$     & 174             & 61               & 4.62            & 180                & 2.348                            & 1323   & 7434              & 40628                  & 5.47    & 0.95      & 1000  & 2.25        \\
MgSiO$_3$    & 245             & 184              & 4.11            & 100                & 2.007                            & 1834   & 10923             & 48140                  & 4.41    & 1.40      & 350   & 0.54        \\
ZrSiO$_4$    & 228             & 109              & 4.675           & 183                & 2.213                            & 1813   & 8936              & 52475                  & 5.87    & 1.03      & 1673  & 2.81        \\
Mg$_2$SiO$_4$   & 129             & 81               & 3.221           & 141                & 2.182                            & 2163   & 8578              & 61728                  & 7.20    & 0.55      & 1473  & 1.45        \\
MgAl$_2$O$_4$   & 197             & 108              & 3.578           & 142                & 2.112                            & 2378   & 9762              & 65681                  & 6.73    & 1.00      & 1500  & 2.06        \\
CaMgSi$_2$O$_6$ & 114             & 65               & 3.31            & 217                & 2.216                            & 1665   & 7786              & 48269                  & 6.20    & 0.74      & 900   & 1.44        \\
LiAlSi$_2$O$_6$ & 123             & 72               & 3.19            & 186                & 2.131                            & 1703   & 8286              & 47479                  & 5.73    & 0.69      & 900   & 1.18        \\
Gd$_3$Ga$_5$O$_{12}$ & 168             & 86               & 7.08            & 1012               & 2.281                            & 2098   & 6319              & 62598                  & 9.91    & 3.00      & 290   & 2.85        \\
Y$_3$Al$_5$O$_{12}$  & 185             & 111              & 4.55            & 594                & 2.213                            & 2223   & 8555              & 64353                  & 7.52    & 4.00      & 300   & 2.14        \\
Al$_6$Si$_2$O$_{13}$ & 172             & 89               & 3.16            & 426                & 2.201                            & 2023   & 9591              & 58238                  & 6.07    & 0.97      & 1273  & 1.75       
\end{longtable}
\normalsize

References for elastic moduli, densities, thermal transport and specific heat:
Si \cite{SiGe, melting,PhysRevLett.120.125901},
Ge \cite{SiGe, melting,semi-kappa, Ge-C},
Diamond \cite{elasticbook, dia1, dia2},
GaP \cite{semi-elas, semi-kappa, semi-C},
GaAs \cite{semi-elas,melting, semi-kappa, semi-C},
GaSb \cite{semi-elas, semi-kappa, semi-C},
InAs \cite{semi-elas, melting, semi-kappa, semi-C},
InP \cite{semi-elas, semi-kappa, semi-C},
InSb \cite{semi-elas, semi-kappa, semi-C},
AlSb \cite{semi-elas, melting, semi-kappa,AlSb-C},
AlN \cite{AlN-elas, Hofmeister2014},
ZnS \cite{elasticbook, ZnS-dif},
ZnSe \cite{ZnSe-elas, ZnSe-kappa,ZnSe-C},
ZnO \cite{elasticbook, oxide-kappa,book-C},
BeO \cite{elasticbook, Hofmeister2014},
NaBr \cite{hof-simple},
LiF \cite{hof-simple},
KCl \cite{elasticbook, hof-simple},
KBr \cite{hof-simple},
RbI \cite{hof-simple},
RbBr \cite{hof-simple},
RbCl \cite{hof-simple},
SrO \cite{elasticbook},
KI \cite{hof-simple},
AgCl \cite{hof-simple},
AgBr \cite{hof-simple},
NaCl \cite{elasticbook, hof-simple},
NaI \cite{hof-simple},
PbS \cite{lead-elas,melting,leadD},
PbSe \cite{lead-elas,melting,leadD},
PbTe \cite{lead-elas,melting,leadD},
MgO \cite{elasticbook,Hofmeister2014},
CaO \cite{elasticbook, oxide-kappa, book-C},
NiO \cite{elasticbook, oxide-kappa, NiO-C},
MnO \cite{elasticbook},
CoO \cite{elasticbook},
ThO$_2$ \cite{lang1958properties, oxide-kappa,book-C},
TiO$_2$ \cite{elasticbook, oxide-kappa, book-C},
UO$_2$ \cite{elasticbook, oxide-kappa, book-C},
SiO$_2$ \cite{elasticbook, book-kappa, book-C},
CaF$_2$ \cite{elasticbook, floride},
SrF$_2$ \cite{SrF2-elas, floride},
BaF$_2$ \cite{elasticbook, floride},
Fe$_2$O$_3$ \cite{elasticbook, Hofmeister2014},
Al$_2$O$_3$ \cite{elasticbook, Hofmeister2014},
Y$_2$O$_3$ \cite{Y2O-elas,lasers},
CaSiO$_3$ \cite{anderson1989theory},
SrTiO$_3$ \cite{elasticperov, hof-perov},
YAlO$_3$ \cite{elasticperov, hof-perov},
BaTiO$_3$ \cite{elasticperov, hof-perov},
KMgF$_3$ \cite{elasticperov},
KZnF$_3$ \cite{elasticperov},
RbMnF$_3$ \cite{elasticperov},
RbCoF$_3$ \cite{elasticperov},
CaTiO$_3$ \cite{elasticperov, hof-perov},
EuAlO$_3$ \cite{elasticperov},
LaAlO$_3$ \cite{LaAl-elas,hof-perov},
KNbO$_3$ \cite{KNb-elas,hof-perov},
MgSiO$_3$ \cite{elasticperov,MgSiO3},
ZrSiO$_4$ \cite{elasticbook, oxide-kappa,ZrSiO4-C},
Mg$_2$SiO$_4$ \cite{elasticbook, oxide-kappa,fosterite},
MgAl$_2$O$_4$ \cite{elasticbook,MgAl2O4},
CaMgSi$_2$O$_6$ \cite{elasticbook, hof-complicated},
LiAlSi$_2$O$_6$ \cite{elasticbook, hof-complicated},
Gd$_3$Ga$_5$O$_{12}$ \cite{garnet-elas,lasers},
Y$_3$Al$_5$O$_{12}$ \cite{garnet-elas,lasers},
Al$_6$Si$_2$O$_{13}$ \cite{mullite-elas, oxide-kappa, mullite}. Where explicit values were not given, we have obtained the bulk $K$ and shear $G$ moduli from the elastic constants. For example, for isotropic
materials $K = (c_{11} + 2 c_{12})/3$ and $G = (c_{11} - c_{12})/2$.

Melting points are mostly from \cite{melting}, where we also consulted \cite{patnaik}. Perovskite melting points mostly from \cite{elasticperov} and \cite{hof-perov}. Some complex material melting points are from \cite{themo-complex}. The garnet melting points are from \cite{melting-garnet}. Throughout, `melting' can also refer to sublimation or dissociation. The interpretation of melting at atmospheric pressure is subtle for diamond and MgSiO$_3$. The values used should be thought of as natural energy scales for the materials. The occasional significant discrepancy that arises between the above sources can be attributed to typos or to distinguishing between melting and dissociation.

\end{document}